\documentclass[aps,prx,showpacs,floatfix,twocolumn,superscriptaddress,10pt]{revtex4-1}

\usepackage[T1]{fontenc}

\usepackage{bm}				
\usepackage{amsmath}
\usepackage{amssymb}
\usepackage{latexsym}
\usepackage{amsfonts}
\usepackage{graphicx}
\usepackage{color}
\usepackage{soul} 

\newcommand{\ket}[1]{\vert{#1}\rangle} 
\newcommand{\bra}[1]{\langle{#1}\vert}

\DeclareMathOperator{\Tr}{Tr}

\newcommand{\be}{\begin{equation}}
\newcommand{\ee}{\end{equation}}

\newcommand{\la}{\langle}
\newcommand{\ra}{\rangle}

\newcommand{\ben}{\begin{eqnarray}}
\newcommand{\een}{\end{eqnarray}}

\newcommand{\cH}{{\cal H}}
\newcommand{\cS}{{\cal S}}
\newcommand{\cB}{{\cal B}}
\newcommand{\cW}{{\cal W}}
\newcommand{\cWl}{{\cal W}_{\lambda}}
\newcommand{\oa}{\omega_{\alpha\beta\nu}(\lambda)}
\newcommand{\loa}{\tilde{\omega}_{\alpha\beta\nu}(\lambda)}

\newcommand{\cBH}{{\cal B}({\cal H})}

\newcommand{\cBa}{\cB_{\alpha\beta}}
\newcommand{\cBaa}{\cB_{\alpha\alpha}}
\newcommand{\cHa}{\cH_{\alpha}}
\newcommand{\PP}{\text{P}}

\newcommand{\ii}{\text{i}}

\begin{document}

\title{Symmetry and the thermodynamics of currents in open quantum systems}

\author{Daniel Manzano}
\email{dmanzano@mit.edu}
\affiliation{Department of Chemistry, Massachusetts Institute of Technology, Cambridge, Massachusetts 02139, USA}
\affiliation{Engineering Product Development, Singapore University of Technology and Design, 20 Dover Drive 138643, Singapore }
\affiliation{Institute for Theoretical Physics, University of Innsbruck, Innsbruck 6020, Austria}
\affiliation{Institute Carlos I of Theoretical and Computational Physics, University of Granada, 18071 Granada, Spain}

\author{Pablo I. Hurtado}
\email{phurtado@onsager.ugr.es}
\affiliation{Institute Carlos I of Theoretical and Computational Physics, University of Granada, 18071 Granada, Spain}
\affiliation{Departamento de Electromagnetismo y F\'isica de la Materia, University of Granada, 18071 Granada, Spain}

\date{\today}


\begin{abstract}
Symmetry is a powerful concept in physics, and its recent application to understand nonequilibrium behavior is providing deep insights and groundbreaking exact results. Here we show how to harness symmetry to control transport and statistics in open quantum systems. Such control is enabled by a first-order-type dynamic phase transition in current statistics {and the associated} coexistence of different transport channels (or nonequilibrium steady states) classified by symmetry. Microreversibility then ensues, via the Gallavotti-Cohen fluctuation theorem, a twin dynamic phase transition for rare current fluctuations. {Interestingly, the symmetry present in the initial state is spontaneously broken at the fluctuating level, where the quantum system selects the symmetry sector that maximally facilitates a given fluctuation.} We illustrate these results in a qubit network model {motivated by the problem} of coherent energy harvesting in photosynthetic complexes, {and introduce the concept of a symmetry-controlled quantum thermal switch, suggesting} symmetry-based design strategies for quantum devices with controllable transport properties.
\end{abstract}

\maketitle

\section{Introduction}
The onset of modern nanotechnologies and the outstanding experimental control of ultracold atoms and trapped ions have just opened the possibility to engineer devices at mesoscopic scales with novel properties and promising technological applications \cite{examp}. Hallmarks of these systems are the importance of quantum effects to understand their dynamics, and the unavoidable interaction with a decohering environment, so the natural framework to describe their properties is the theory of open quantum systems \cite{Petruccione}. Due to their mesoscopic size, their physics is typically dominated by large fluctuations that determine their function and response. In addition, these devices usually operate under nonequilibrium conditions, so a full understanding of their physics is only possible by analyzing their nonequilibrium fluctuating behavior, with particular emphasis on the statistics of currents, a key observable out of equilibrium. The natural language for this program is the theory of large deviations 
or full-counting statistics \cite{Touchette,Mukamel}, recently extended to the realm of open quantum systems \cite{Garrahan}, with the current large deviation function (LDF) measuring the the probability of current fluctuations as central object in the theory. Advancing this line of research is both of fundamental and practical importance. On one hand, the current LDF plays in nonequilibrium a role equivalent to the equilibrium free energy, governing the \emph{thermodynamics of currents} and hence the transport and collective behavior out of equilibrium \cite{qLDF1,qLDF2,PabloJSP}. On the other hand, {as we show in this paper,} a detailed understanding of the transport and fluctuating properties of open quantum systems {and the role of symmetry} is {helpful} to devise optimal quantum control strategies in open systems \cite{opensim,abu}, dissipation-engineered state preparation \cite{Naturecomp2,Kraus,Wineland} and dissipation-driven quantum computation \cite{Naturecomp}, all important for emerging 
technological applications.

Despite the increasing interest and efforts along these lines, understanding the physics of nonequilibrium systems, classical or quantum, is remarkably challenging. This is due to the difficulty in combining statistics and dynamics, which always plays a key role out of equilibrium \cite{qLDF1}. Most prominent among the few general results in nonequilibrium physics are the different fluctuation theorems \cite{GC,ECM,LS,K,IFR,Mukamel}, which strongly constraint the probability distributions of fluctuations far from equilibrium. These theorems are different expressions of a symmetry, the time reversibility of microscopic dynamics, at the {meso}scopic, irreversible level, illustrating the power of symmetry as a tool to obtain new insights into nonequilibrium behavior. Symmetry ideas \cite{SymGross} have already proved useful to study transport in quantum systems. For instance, geometric symmetries of the Hamiltonian trigger anomalous collective quantum effects like superradiance (enhanced relaxation rate) \
cite{srad} and supertransfer (enhanced exciton transfer rate and diffusion length) \cite{strans}. Another examples concern the strong constraints imposed by symmetries of the reduced density matrix on the nonequilibrium steady states of open quantum systems \cite{Livi}, or the enhancement of quantum transport by time-reversal symmetry breaking found in continuous-time quantum walks \cite{natureTRSB}. In addition, symmetry principles have been recently used to devise optimal quantum control strategies \cite{control}.

\begin{figure}[b]
\vspace{-0.5cm}
\includegraphics[width=8cm]{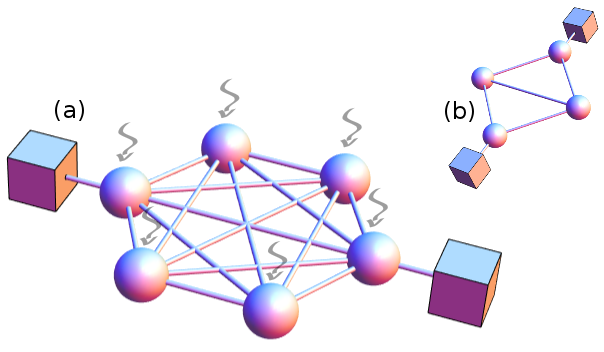}
\caption{\small (Color online) (a) Fully-connected network of 6 qubits (spheres) in contact with two thermal baths (boxes) and possibly subject to dephasing noise (wavy arrows). Symmetries correspond to permutations of bulk pairs. {(b) Sketch of a symmetry-controlled quantum thermal switch (see text).}}
\label{fig1}
\end{figure}

Inspired by these illuminating results, we explore in this paper the consequences of a symmetry for the transport properties and the current statistics of open quantum systems of the Lindblad form \cite{Petruccione,Lindblad}. For these systems, it has been recently shown that the existence of a symmetry implies the emergence of multiple nonequilibrium steady states classified via the symmetry spectrum  \cite{Prosen} (see also \cite{Baumgartner}). As we show below, this coexistence of different transport channels is {associated with} a general first-order-type dynamic phase transition in the statistics of current fluctuations, which shows up as a kink in the cumulant generating function of the current distribution, or equivalently as a non-convex regime in {the current LDF}, {and can be used to control transport and fluctuations in open quantum systems by tailoring the symmetry-protected information present in the initial state}. {Interestingly,} the original symmetry of the open quantum system is 
{spontaneously} broken at the fluctuating level, where the system naturally selects a particular symmetry subspace with maximal or minimal current to facilitate a given current fluctuation. Using the time-reversibility of microscopic dynamics, we further prove that this instability is accompanied by a \emph{twin dynamic phase transition} for rare, reversed current fluctuations. Remarkably, the twin dynamic phase transitions are a purely nonequilibrium effect, disappearing in equilibrium. As an example of the power of this method, we study current statistics in fully-connected open networks of qubits, see Fig. \ref{fig1}.a, a model of coherent energy harvesting where symmetry-controlled transport and twin dynamic phase transitions are clearly demonstrated.The understanding of transport in quantum networks is currently under intense investigation since recent experiments suggested coherent energy transfer in the Fenna-Matthews-Olson complex of green sulphur bacteria \cite{FMO}, even at room temperature. Our results show how symmetry principles can be used to unveil exact and general results in nonequilibrium open quantum systems, and suggest novel design strategies based on symmetry ideas for quantum devices with controllable transport properties. {In fact, using this approach we propose a novel design for a \emph{symmetry-controlled quantum thermal switch}, see Fig. \ref{fig1}.b, i.e. a quantum qubit device where the heat current between a hot and a cold reservoirs can be completely blocked, modulated or turned on by preparing the symmetry of the initial state.}

\section{Symmetry and the thermodynamics of currents} 

We consider an open quantum system weakly coupled to its environment. The state of such system is described at any time by a (reduced) density matrix $\rho$, a trace-one operator in the space $\cBH$ of bounded operators acting on the system's Hilbert space $\cH$, which we assume of finite dimension $D$. The space $\cBH$ is itself a Hilbert space once supplemented with an appropriate inner product, the Hilbert-Schmidt product $(\sigma,\rho)\equiv \Tr(\sigma^{\dagger} \rho)$, $\forall \sigma,\, \rho \in \cBH$. For Markovian open quantum systems, the density matrix evolves according to the well-known Lindblad master equation \cite{Petruccione,Lindblad}
\be
\dot{\rho}= -\ii [H,\rho] +  \sum_m \left(L_m\, \rho\, L^\dagger_m - \frac{1}{2}\{L^\dagger_m L_m,\rho\}\right) \equiv  \cW \rho \, ,
\label{Lindblad}
\ee
where $H$ is the system Hamiltonian, $[A,B]=AB-BA$ and $\{A,B\}=AB+BA$, and $L_m\in\cBH$ are the so-called Lindblad operators \cite{Petruccione}. {Note also that we employ units of  $\hbar = 1$ throughout the paper}. This equation defines the evolution \emph{superoperator} $\cW$ --a $D^2\times D^2$ matrix acting on $\cBH$-- and describes the \emph{coherent} evolution of an open quantum system (as captured by the first term in the rhs), punctuated by weak, \emph{decohering} interactions with a fast-evolving environment (modeled by the Lindblad operators). Eq. (\ref{Lindblad}) is the most general dynamical law for the reduced density matrix of an open Markovian quantum system which preserves normalization and is completely positive \cite{Petruccione}. We are interested in Lindblad operators describing most common physical situations, namely (i) coupling to \emph{different} reservoirs (of energy, magnetization, etc.), which locally inject and extract excitations at constant rate, or (ii) the effect of environmental dephasing noise which causes local decoherence and thus classical 
behavior \cite{Petruccione}. In this way, Eq. (\ref{Lindblad}) describes all sorts of nonequilibrium situations driven by external gradients and noise sources. Steady states now correspond to the null fixed points of the Lindblad superoperator, $\cW \rho_{\text{st}} = 0$. In a recent theorem \cite{Prosen}, Bu\v ca and Prosen have shown that an open quantum system of this sort with a \emph{strong symmetry}, i.e. an unitary operator $S\in \cBH$ that simultaneously commutes with the Hamiltonian and all Lindblad operators, $[S,H]=0=[S,L_m]$ $\forall m$, will necessarily have multiple, degenerate (nonequilibrium) steady states, hereafter NESSs, which can be indexed by the symmetry $n_s$ distinct eigenvalues. In fact, using the symmetry spectrum defined via $S\ket{\psi_{\alpha}^{(k)}}=\exp(i\theta_{\alpha}) \ket{\psi_{\alpha}^{(k)}}$, with eigenvectors $\ket{\psi_{\alpha}^{(k)}}\in \cH$, $\alpha\in[1,n_s]$, $k\in[1,d_{\alpha}]$ and $d_{\alpha}$ the dimension of each eigenspace, we may introduce spectral decompositions of both the Hilbert space $\cH=\bigoplus_{\alpha} \cH_{\alpha}$, with $\cH_{\alpha}=\{\ket{\psi_{\alpha}^{(k)}},k\in [1,d_{\alpha}]\}$, and the operator Hilbert space $\cBH=\bigoplus_{\alpha\beta} \cB_{\alpha\beta}$, with $\cB_{\alpha\beta}=\{\ket{\psi_{\alpha}^{(n)}} \bra{\psi_{\beta}^{(m)}}: n\in [1,d_{\alpha}],m\in [1,d_{\beta}]\}$ and dimension $d_{\alpha\beta}=d_{\alpha} d_{\beta}$. By defining the left and right adjoint symmetry superoperators $\cS_{L,R}$ as $\cS_L \eta = S \eta$ and $\cS_R \eta = \eta S^{\dagger}$ $\forall \eta\in \cBH$, it is clear that the subspaces $\cB_{\alpha \beta}$ are the joint eigenspaces of both $\cS_L$ and $\cS_R$. It is then an exercise to show, using the commutation relations above defining the strong symmetry $S$, that the subspaces $\cBa$ remain invariant under the flow $\cW$, i.e. $\cW \cBa \subset \cBa$, and hence each subspace contains at least one well-defined and different fixed point of the dynamics \cite{Prosen,Albert,note1}. By noting that trace-
one, \emph{physical} density matrices can only live in diagonal subspaces $\cBaa$ due to the orthogonality between the different $\cHa$, we obtain at least $n_s$ distinct NESSs, one for each $\cBaa$, which can be labeled according to the symmetry eigenvalues, i.e. for any normalized $\rho_{\alpha}(0)\in \cBaa$ we have 
$\rho_{\alpha}^{\scriptscriptstyle\text{NESS}}\equiv \lim_{t\to \infty} \exp{(+t\cW)}\rho_{\alpha}(0)\in \cBaa$, and a continuum of possible linear combinations of these NESSs. 
The different NESSs $\rho_{\alpha}^{\scriptscriptstyle\text{NESS}}$ can be further degenerate according to the Evans theorem \cite{Evans}, though we assume here for simplicity that $\rho_{\alpha}^{\scriptscriptstyle\text{NESS}}$ are unique for each $\alpha$. Interestingly, one-dimensional symmetry eigenspaces $\ket{\psi_{\alpha}} \bra{\psi_{\alpha}}$ will be mapped onto themselves by the dynamics $\cW$, thus defining decoherence-free, \emph{dark} states  important e.g. in quantum computing to protect quantum states from relaxation \cite{Prosen,Albert,Blume}. 

Our aim now is to study the implications of such strong symmetry for the statistics of the current flowing through a given reservoir, a main observable out of equilibrium \cite{Mukamel,qLDF1,qLDF2,PabloJSP,Prosen2,znidaric1,Dhar,Maes,Gaspard}. For that, we first introduce the reduced density matrix $\rho_Q(t)$, which is the projection of the full density matrix to the space of $Q$ events, being $Q$ the total (energy, spin, exciton, \ldots) current flowing from a reservoir to the system in a time $t$. This current can be appropriately defined in the quantum realm via the \emph{unraveling} of the master equation (\ref{Lindblad}) \cite{Maes}. The probability of observing a given current fluctuation, typical or rare, is thus $\PP_t(Q)=\Tr[\rho_Q(t)]$, and scales in a large deviation form for long times, $\PP_t(Q)\asymp \exp[+tG(Q/t)]$, where $G(q)\leq 0$ is the current large deviation function \cite{Touchette,Mukamel,Garrahan,qLDF1,qLDF2,PabloJSP}. This scaling shows that the probability of observing a significant current fluctuation away from its average is exponentially small in time. As usual in statistical physics, it is difficult to work with a global constraint (think for instance on the microcanonical ensemble), and the problem becomes simpler after an appropriate change of \emph{ensemble}. With this idea in mind, we introduce the Laplace transform $\rho_{\lambda}(t)=\sum_Q \rho_{Q}(t) \exp(-\lambda Q)$ with $\lambda$ a \emph{counting field} conjugated to the current, such that $Z_{\lambda}(t)\equiv \Tr[\rho_{\lambda}(t)]$ corresponds to the moment generating function of the current distribution, which also obeys a large deviation principle of the form $Z_{\lambda}(t)\asymp \exp[+t\mu(\lambda)]$ for long times. Here $\mu(\lambda)=\max_q[G(q)-\lambda q]$ is the Legendre transform of the current LDF, in a way equivalent to the thermodynamic relation between the canonical and microcanonical potentials \cite{Touchette,Mukamel,Garrahan,qLDF1,qLDF2,PabloJSP}. Interestingly, $\rho_Q(t)$ obeys a complex hierarchy of equations which is however disentangled by the Laplace transform \cite{Mukamel,Zoller}, yielding a closed evolution equation for $\rho_{\lambda}(t)$
\ben
\dot{\rho}_{\lambda}(t) &=&  - \ii [H,\rho_{\lambda}] +  \text{e}^{-\lambda} L_1 \rho_{\lambda} {L_1}^\dagger  +  \text{e}^{+\lambda} L_2 \rho_{\lambda} {L_2}^\dagger \label{rhol}  \\
&+& \sum_{m\neq 1,2} L_m \rho_{\lambda} L^\dagger_m -  \frac{1}{2}\sum_{m} \{L^\dagger_m L_m,\rho_{\lambda}\} \equiv \cWl \rho_{\lambda} , \nonumber
\een
where we assume without loss of generality that $L_1$ and $L_2$ are respectively the Lindblad operators responsible of the injection and extraction of excitations through the reservoir of interest. This defines a deformed superoperator $\cWl$ which no longer preserves the trace, and whose spectral properties determine the thermodynamics of currents in the system at hand. 

The existence of a strong symmetry implies that the symmetry superoperators $\cS_{L,R}$ and $\cWl$ all commute, so there exists a complete biorthogonal set of common left ($\loa$) and right ($\oa$) eigenfunctions in $\cBH$, linking eigenvalues of $\cWl$ to particular symmetry eigenspaces, such that $\cS_L\oa=\text{e}^{i\theta_{\alpha}}\oa$, $\cS_R\oa=\text{e}^{-i\theta_{\beta}}\oa$, and $\cWl\oa=\mu_{\nu}(\lambda)\oa$ (similarly for left eigenfunctions). Note that, due to orthogonality of symmetry eigenspaces, $\Tr[\oa]\propto \delta_{\alpha\beta}$, and we introduce the normalization $\Tr[\omega_{\alpha\alpha\nu}(\lambda)]=1$ for simplicity. The solution to Eq. (\ref{rhol}) can be formally written as $\rho_{\lambda}(t)=\exp(+t \cWl)\rho(0)$, so a spectral decomposition of the initial density matrix in terms of the common biorthogonal basis yields $Z_{\lambda}(t)=\sum_{\alpha\nu}\text{e}^{+t\mu_\nu(\lambda)} \left(\tilde{\omega}_{\alpha\alpha\nu}(\lambda),\rho(0)\right)$. For long times
\be
Z_{\lambda}(t) \xrightarrow{t\to\infty} \text{e}^{+t\mu_0^{(\alpha_0)}(\lambda)} \left(\tilde{\omega}_{\alpha_0\alpha_0 0}(\lambda),\rho(0)\right) \, ,
\label{zsum1}
\ee
where $\mu_0^{(\alpha_0)}(\lambda)$ is the eigenvalue of $\cWl$ with largest real part and symmetry {index} $\alpha_0$ among all symmetry diagonal eigenspaces $\cBaa$ with nonzero projection on the initial $\rho(0)$. In this way, this eigenvalue defines the Legendre transform of the current LDF, $\mu(\lambda)\equiv\mu_0^{(\alpha_0)}(\lambda)$, see above. Interestingly, the long time limit in Eq. (\ref{zsum1}) selects a particular symmetry eigenspace $\alpha_0$ (assumed here unique in order not to clutter our notation; this is however unimportant for our conclusions below), effectively breaking at the fluctuating level the original symmetry of our open quantum system. As we show below, distinct symmetry eigenspaces may dominate different fluctuation regimes, separated by first-order-type dynamic phase transitions. Note that a different type of spontaneous symmetry breaking scenario at the fluctuating level has been recently reported in classical diffusive systems \cite{qLDF1,PabloJSP,PabloSSB,BD2,CarlosSSB}. 

\begin{figure*}
\includegraphics[width=12cm]{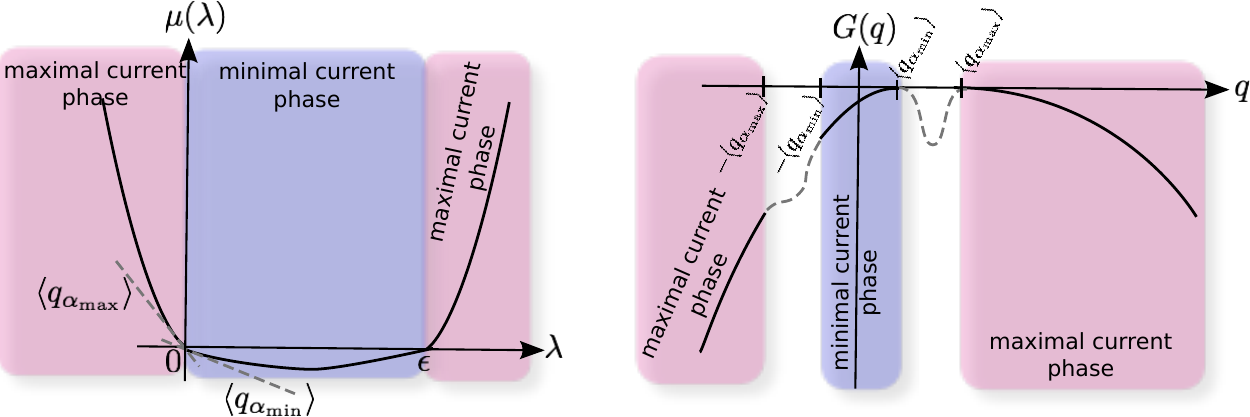}
\caption{\small (Color online) Sketch of the twin dynamic phase transitions in the current statistics of an open quantum system with a symmetry, as appears for the current cumulant generating function $\mu(\lambda)$ (left), and the associated current large deviation function $G(q)$ (right). Notice the twin kinks in $\mu(\lambda)$ and the corresponding non-convex regimes in $G(q)$ \cite{nonconvex}.}
\label{fig2}
\end{figure*}

The previous arguments also show how it is possible to control both the statistics of the current and the average transport properties of an open quantum system by playing with the symmetry decomposition of the initial state $\rho(0)$, which in turn controls the amplitude of the scaling in Eq. (\ref{zsum1}) {(for a discussion of this amplitude in a classical context, see} \cite{Dhar}). This is most evident by studying the average current, defined as $\la q \ra = \lim_{t\to\infty} \frac{1}{t} \partial_{\lambda} \ln Z_{\lambda}(t)\vert_{\lambda=0}$. Using again the previous spectral decomposition, it is easy to show that
\be
\la q \ra = \frac{\sum_{\alpha} \la q_{\alpha}\ra (\tilde{\rho}_{\alpha}^{\scriptscriptstyle\text{NESS}},\rho(0))}{\sum_{\alpha} (\tilde{\rho}_{\alpha}^{\scriptscriptstyle\text{NESS}},\rho(0))} \,
\label{qmed}
\ee
where {$\la q_{\alpha}\ra = -\partial_{\lambda} \mu_0^{(\alpha)}(\lambda)\vert_{\lambda=0}=\Tr[L_2\rho_{\alpha}^{\scriptscriptstyle\text{NESS}}L_2^\dagger]-\Tr[L_1\rho_{\alpha}^{\scriptscriptstyle\text{NESS}}L_1^\dagger]$} is the average current of the NESS $\rho_{\alpha}^{\scriptscriptstyle\text{NESS}}\in\cBaa$. To derive Eq. (\ref{qmed}) we have used that $\cW_{\lambda=0}=\cW$, whose largest eigenvalue within each symmetry eigenspace $\cBaa$ is necessarily 0 \cite{Albert,note1}, with associated normalized right eigenfunction $\omega_{\alpha\alpha 0}(\lambda=0)=\rho_{\alpha}^{\scriptscriptstyle\text{NESS}}$ and dual $\tilde{\rho}_{\alpha}^{\scriptscriptstyle\text{NESS}}${, see Appendix A}. Nonequilibrium steady states $\rho_{\alpha}^{\scriptscriptstyle\text{NESS}}$ with different $\alpha$ will typically have different average currents $\la q_{\alpha}\ra$, so the manipulation of the projections $(\tilde{\rho}_{\alpha}^{\scriptscriptstyle\text{NESS}},\rho(0))$ by adequately preparing the symmetry of the 
initial state will lead to symmetry-controlled transport properties. We show below {several} examples of this control mechanism.

{Remarkably,} the existence of a symmetry under nonequilibrium conditions implies nonanalyticities in the LDF $\mu(\lambda)$ which can be interpreted as dynamical phase transitions {separating regimes where the original symmetry is spontaneously broken in different ways.} {To show this}, we first note that for $|\lambda|\ll 1$ the leading eigenvalue of $\cWl$ with symmetry {index} $\alpha$ can be expanded as $\mu_0^{(\alpha)}(\lambda)\approx \mu_0^{(\alpha)}(0) + \lambda (\partial_{\lambda} \mu_0^{(\alpha)}(\lambda))\vert_{\lambda=0}=-\lambda \la q_{\alpha}\ra$. Therefore, by using that $\mu(\lambda)=\max_{\alpha}[\mu_0^{(\alpha)}(\lambda)]$, the maximum taken over the symmetry eigenspaces with nonzero overlap with $\rho(0)$, we arrive at
\be
\mu(\lambda) \underset{|\lambda|\ll 1}{=} \left\{
\begin{array}{l l}
  +|\lambda| \la q_{\alpha_{\text{max}}} \ra  & \quad \text{for} \, \, \lambda \lesssim 0\\
  -|\lambda| \la q_{\alpha_{\text{min}}} \ra & \quad \text{for} \, \, \lambda \gtrsim 0
  \end{array} \right. \, ,
\ee
where $\alpha_{\text{max}}$ ($\alpha_{\text{min}}$) denotes the symmetry eigenspace with maximal (minimal) average current $\la q_{\alpha_{\text{max}}} \ra$ ($\la q_{\alpha_{\text{min}}} \ra$) among those with nonzero overlap with $\rho(0)$. Therefore the LDF $\mu(\lambda)$ exhibits a kink at $\lambda=0$, characterized by a finite, discontinuous jump in the dynamic order parameter $q(\lambda)\equiv -\mu'(\lambda)$ at $\lambda=0$ of magnitude $\Delta q_0=\la q_{\alpha_{\text{max}}} \ra - \la q_{\alpha_{\text{min}}} \ra$, a behavior reminiscent of first order phase transitions \cite{Garrahan}. Furthermore, if the original evolution superoperator $\cW$ is microreversible (i.e. obeys a detailed balance condition)  \cite{Mallick,Agarwal,Gaspard}, the system of interest will obey the Gallavotti-Cohen fluctuation theorem for currents, which links the probability of a current fluctuation with its time-reversal event \cite{GC,ECM,LS,K,IFR}. This fluctuation theorem can be stated as $\mu(\lambda)=\mu(\epsilon-\lambda)$ for the Legendre transform of the LDF, where $\epsilon$ is a constant related to the rate of entropy production in the system. In this way, we see that the kink in $\mu(\lambda)$ observed at $\lambda=0$ is reproduced at $\lambda=\epsilon$, where a \emph{twin dynamic phase transition} emerges, see Fig. \ref{fig2}. By inverse Legendre transforming $\mu(\lambda)$ to obtain the current LDF $G(q)=\max_{\lambda}[\mu(\lambda) + q\lambda]$, it is straightforward to show \cite{Touchette} that the twin kinks in $\mu(\lambda)$ corresponds to two different current intervals, $|q|\in[|\la q_{\alpha_{\text{min}}} \ra|,|\la q_{\alpha_{\text{max}}} \ra|]$, related by time-reversibility or $q\leftrightarrow -q$, where $G(q)$ is {non-convex}, see Fig. \ref{fig2}. This corresponds to a multimodal current distribution $\PP_t(Q=qt)$, with several peaks  reflecting the \emph{coexistence} of multiple transport channels, each one associated with a different NESS in our open quantum system with a strong symmetry \cite{Prosen}. Remarkably, the symmetry is broken at the fluctuating level, where the quantum system selects a symmetry {sector} that maximally facilitates a given current fluctuation: the statistics during a current fluctuation with $|q|>|\la q_{\alpha_{\text{max}}} \ra|$ is dominated by the symmetry eigenspace with maximal current ($\alpha_{\text{max}}$), whereas for $|q|<|\la q_{\alpha_{\text{min}}} \ra|$ the minimal current eigenspace ($\alpha_{\text{min}}$) prevails. This is best captured by the effective density matrix $\rho_{\lambda}^{\text{eff}}\equiv \lim_{t\to\infty} \rho_{\lambda}(t)/\Tr[\rho_{\lambda}(t)]=\omega_{\alpha_0\alpha_0 0}(\lambda)$, with $\alpha_0=\alpha_{\text{max}}$ ($\alpha_{\text{min}}$) for $|\lambda-\frac{\epsilon}{2}|>\frac{\epsilon}{2}$ ($|\lambda-\frac{\epsilon}{2}|<\frac{\epsilon}{2}$).

Interestingly, the previous twin dynamic phase transitions in current statistics only happen out of equilibrium, disappearing in equilibrium. In {the latter} case, the average currents for the multiple steady states are zero in all cases, $\la q_{\alpha}\ra =0 \,\, \forall \alpha$, so no symmetry-induced kink appears in $\mu(\lambda)$ at $\lambda=0$ in equilibrium \cite{kink}. Moreover, an expansion for $|\lambda|\ll 1$ of the leading eigenvalues yields to first order $\mu_0^{(\alpha)}(\lambda)\approx  \frac{\lambda^2}{2} (\partial_{\lambda}^2 \mu_0^{(\alpha)}(\lambda))\vert_{\lambda=0}=\frac{\lambda^2}{2} \sigma_{\alpha}^2$, where $\sigma_{\alpha}^2$ is the variance of the current distribution in each steady state, so for equilibrium systems the overall current statistics is dominated by the symmetry eigenspace with maximal variance among those present in the initial $\rho(0)$. Therefore it is still possible to control the statistics of current fluctuations in equilibrium by an adequate preparation of $\rho(0)$, though $G(q)$ is convex around $\la q\ra=0$ and no dynamic phase transitions are expected. 

\begin{figure}
\vspace{-0.6cm}
\includegraphics[width=9cm]{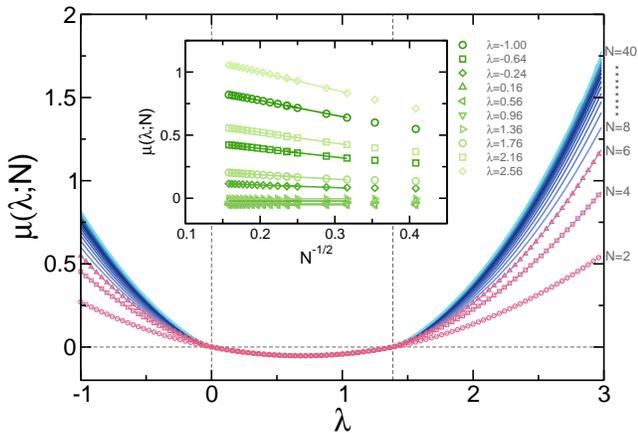}
\vspace{-0.75cm}
\caption{\small (Color online) Main: The current cumulant generating function $\mu(\lambda;N)$ as a function of $\lambda$ for different $N$ {and parameters $a_1=2=b_2$, $b_1=1=a_2$, and $h=1=J$}. Open symbols correspond to the numerical diagonalization of the full deformed Lindblad superoperator $\cWl$, while lines show results after the symmetry-induced dimensional reduction. The vertical dashed lines signal the critical points $\lambda=0,\epsilon$. While no $N$-dependence is observed for $0<\lambda<\epsilon$, a rapid increase with size appears outside this interval, suggesting the emergence of two kinks in $\mu(\lambda;N)$ at $\lambda=0,\epsilon$. Inset: $N$-dependence of $\mu(\lambda;N)$ for different fixed $\lambda$. A clear $N^{-1/2}$ scaling is evident. 
}
\label{fig3}
\end{figure}

\section{Application to open quantum networks} 
The study of energy transport in quantum networks has recently attracted a lot of attention, since empirical evidences of coherent transport {at room temperature} have been found in the the Fenna-Matthews-Olson complex of green sulfur bacteria \cite{FMO}. This complex plays an important role during the photosynthetic process by conducting energy from the antenna {through a heterogeneous chromophore network} to the reaction center, where the photosynthetic reaction takes place. Motivated by this energy harvesting problem, we now proceed to apply the general results {of the previous section} to study transport in open quantum networks \cite{manzano:po13,caruso:pra10}. {These are oversimplified models of quantum transport which have proven extremely useful to understand the functional role of noise and dephasing in enhancing coherent energy transfer.}

We {hence} study homogeneous fully-connected networks of $N$ quantum two-level systems, or qubits, see Fig. \ref{fig1}.a. We focus here on $N$ even for simplicity, though similar results hold for odd $N$. The Hamiltonian is
\be 
H= h \sum_{i=1}^N \sigma_i^+ \sigma_i^- + J \sum_{\substack{i,j=1 \\  j< i}}^N  \left( \sigma_i^+ \sigma_j^- +  \sigma_i^- \sigma_j^+  \right) \, , 
\label{hamil}
\ee 
where $\sigma_i^+$ and $\sigma_i^- $ are the raising and lowering operators acting on qubit $i$, $h$ is the on-site energy, and $J$ represents the coupling strength. {The nonequilibrium, dissipative dynamics of the system is triggered by two Markovian bosonic heat baths that pump and extract excitations in an incoherent way from qubits $1$ and $N$. We will refer to these qubits as  \emph{terminal}, while the remaining qubits form the \emph{bulk}. The full system dynamics, including the incoherent hopping from the baths, can be described by a Markovian master equation (\ref{Lindblad}) \cite{Petruccione} with 4 Lindblad operators, $L_1=\sqrt{a_1}\sigma_1^+$ and $L_2=\sqrt{b_1}\sigma_1^-$ for the first bath, and $L_3=\sqrt{a_N}\sigma_N^+$ and $L_4=\sqrt{b_N}\sigma_N^-$ for the second. The bath constants $a_i$ and $b_i$ account for the excitation pumping and extraction rates at qubit $i$, respectively, and a temperature gradient sets in whenever $a_1 b_2\neq a_2 b_1$. In fact, the external nonequilibrium drive can be quantified by $\epsilon=\ln[a_1 b_2/(a_2 b_1)]$.} 

Similar qubit models, with dipole-dipole interactions, have been also studied in order to analyze quantum Fourier's law \cite{manzano:pre12,znidaric:pre11} and energy transfer in quantum networks, both in the transient \cite{caruso} and steady state regimes \cite{manzano:po13}. {The Hamiltonian (\ref{hamil}) is also related with that of the Lipkin-Meshkov-Glick model}, that was {introduced} in 1965 to describe phase transitions in nuclei \cite{lipkin}. {For a closed system, with no coupling to an external environment, a}n exact solution of this model can be {obtained starting from} Bethe equations \cite{ribeiro}, {though analytical solutions in an open framework are still lacking. We expect our results below on the effect of symmetry on the thermodynamics of currents may help in this effort.}

\begin{figure}
\includegraphics[width=8.5cm]{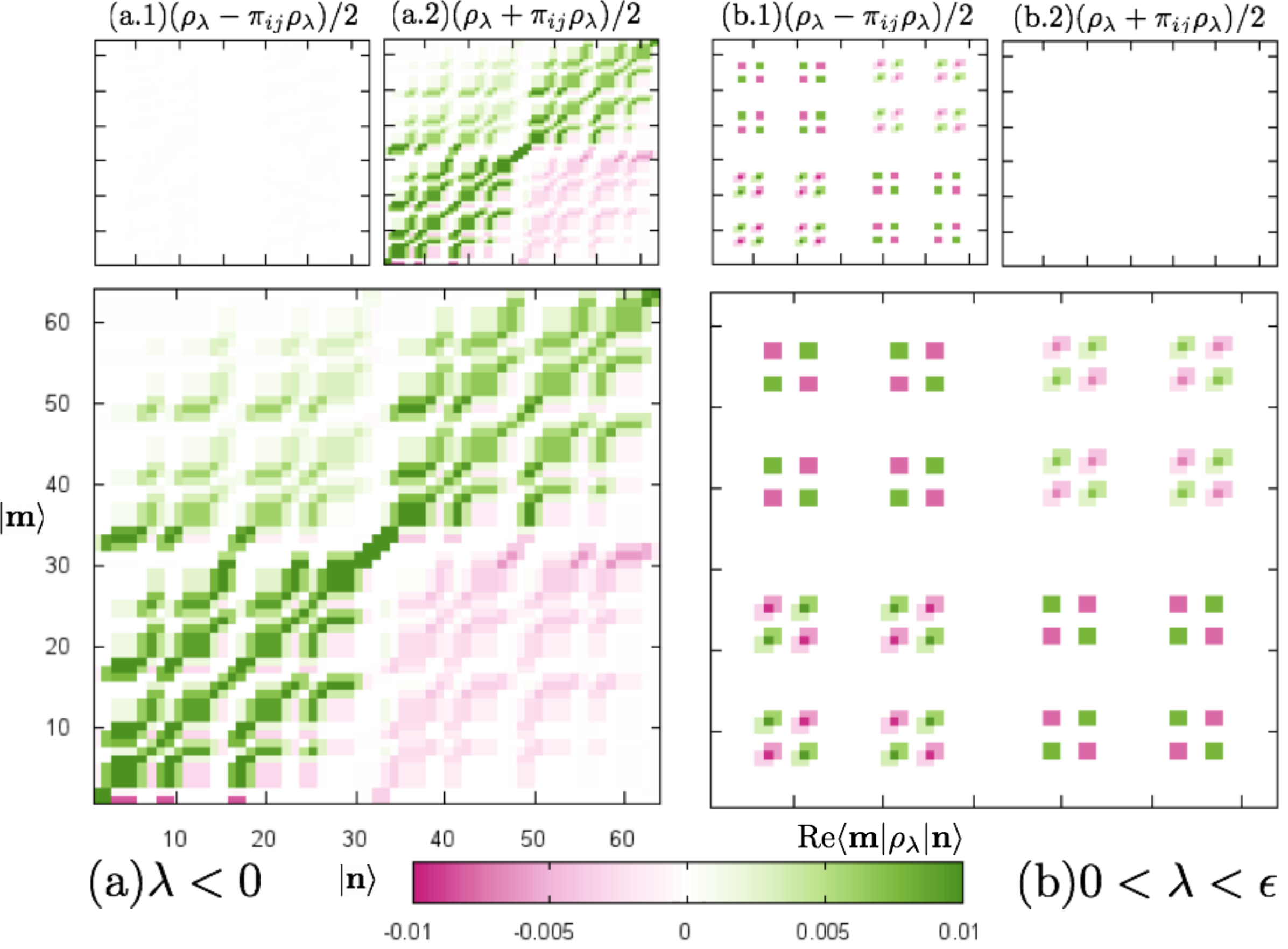}
\vspace{-0.25cm}
\caption{\small (Color online)  Real part of the $N=6$ normalized right eigenmatrix $\omega_{\alpha_0\alpha_0 0}(\lambda)$ associated with the eigenvalue of $\cWl$ with largest real part, for (a) $\lambda=-0.4$ and (b) $\lambda=0.2$. Panels (a.1)-(b.1) and (a.2)-(b.2) show respectively the $(i,j)$-antisymmetrized and -symmetrized eigenmatrices, with $(i,j)$ an arbitrary pair of bulk qubits. For $\lambda<0$ (and $\lambda>\epsilon$) the leading eigenmatrix is completely symmetric, while for $0<\lambda<\epsilon$ it is pair-antisymmetric. 
{System parameters as in Fig. \ref{fig3}.}
}
\label{fig4}
\end{figure}

\begin{figure}[t]
\vspace{-0.5cm}
\includegraphics[width=9cm]{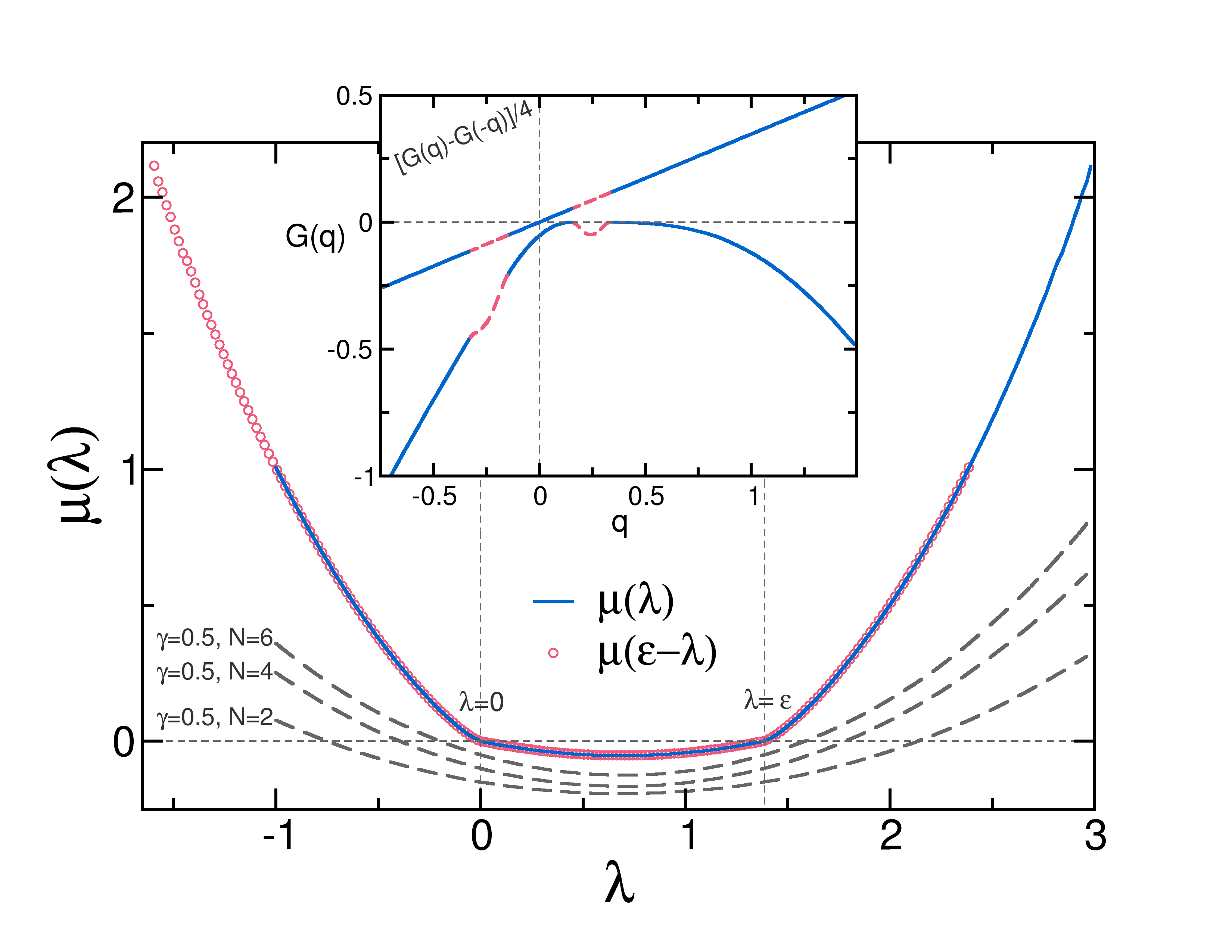}
\vspace{-0.8cm}
\caption{\small (Color online) Estimation of $\mu(\lambda)$ in the thermodynamic limit as obtained from the $N^{-1/2}$ scaling observed in the inset to Fig. \ref{fig3}. Twin kinks at $\lambda=0,\epsilon$ (signaled by thin vertical dashed lines) are apparent, and the current distribution obeys the Gallavotti-Cohen (GC) fluctuation theorem, $\mu(\lambda)=\mu(\epsilon-\lambda)$. Dashed thick lines show $\mu(\lambda)$ measured for networks with dephasing noise ($\gamma=0.5$). Curves have been shifted downward for clarity (in all cases $\mu(0)=0$). Dephasing destroys the permutation symmetry, and the twin dynamic phase transitions disappear. Inset: Asymptotic current LDF obtained from the numerical inverse Legendre transform of $\mu(\lambda)$ in the main panel. Dashed lines sketch the non-convex regimes of $G(q)$ for which $\mu(\lambda)$ offers no information. Again, the GC fluctuation theorem is clearly satisfied, $G(q)-G(-q)=\epsilon q$. 
}
\label{fig5}
\end{figure}

Remarkably, this model exhibits not just one, but multiple strong symmetries in the sense of ref. \cite{Prosen} for $N\ge 4$. In fact, any permutation $\pi_{ij}\in\cBH$ exchanging the state of a pair of bulk qubits $i,j\in[2,N-1]$ leaves invariant the Hamiltonian (\ref{hamil}) \cite{ribeiro}, and obviously commutes with the Lindblad operators $L_m$ $\forall m$ as they only affect terminal qubits. Therefore $[\pi_{ij},H]=0=[\pi_{ij},L_m]$, so we expect the open quantum network to exhibit multiple NESSs classified by the permutations spectrum (thus allowing for symmetry-controlled transport), together with a pair of twin dynamic phase transitions involving a symmetry-breaking event. To see this, we analyzed the spectrum of the deformed superoperator $\cWl$ for this particular model, see Eq. (\ref{rhol}), looking for the leading eigenvalue which defines the current LDF $\mu(\lambda;N)$ and the associated right eigenmatrix. {For simplicity, we focus hereafter on a particular set of parameters, namely $h=1=J$ and $a_1=2=b_2$, $b_1=1=a_2$, corresponding to $\epsilon\approx 1.39$.} Open symbols in Fig. \ref{fig3} show results for $\mu(\lambda;N)$ {in this case} as obtained by numerically diagonalizing $\cWl$ for $N=2,\, 4$ and 6 qubits. Note that $\cWl$ is a $4^N\times4^N$ matrix, an exponential size scaling which prevents us from reaching larger networks with this method (see however below). Interestingly, the measured $\mu(\lambda;N)$ shows no dependence on $N$ for $0<\lambda<\epsilon$, while a rapid increase with size appears outside this interval, $|\lambda-\frac{\epsilon}{2}|>\frac{\epsilon}{2}$. This behavior suggests the presence of two kinks in $\mu(\lambda;N)$ at $\lambda=0,\epsilon$ for $N\ge 4$, where $\partial_{\lambda}\mu(\lambda;N)$ becomes discontinuous. The sharp change of behavior at $\lambda=0,\epsilon$ is most evident when studying the associated leading eigenmatrix. Fig. \ref{fig4} plots the real part of the eigenmatrix in the computational basis measured for $N=6$ for two values of $\lambda$ across the kink at $\lambda=0$. The qualitative difference of the two eigenmatrices is confirmed when studying its behavior under permutations of bulk qubits. In fact, the measured eigenmatrix for $\lambda<0$ (as well as for $\lambda>\epsilon$) is \emph{completely symmetric} under any permutation of bulk qubits, see Figs. \ref{fig4}.a.1-2, while for $0<\lambda<\epsilon$ the resulting eigenmatrix is \emph{antisymmetric by pairs}, i.e. with non-overlapping pairs of bulk qubits in antisymmetric, singlet state, see Figs. \ref{fig4}.b.1-2 (note that this regime is degenerate for $N>4$ as bulk qubits can be partitioned by pairs in different ways). This confirms the existence of a pair of twin symmetry-breaking dynamic phase transitions happening at $\lambda=0,\epsilon$ (equivalent results hold for $N=4$). For large current fluctuations such that $|\lambda-\frac{\epsilon}{2}|>\frac{\epsilon}{2}$, the quantum network selects the symmetry subspace with maximal current, which corresponds to the totally symmetric subspace. This sort of \emph{bosonic transport regime} can be understood phenomenologically by noting that a totally symmetric bulk can absorb a maximal number of excitations from the terminal qubit, hence leaving it free to receive further excitations from the reservoir and thus maximizing the current flowing through the system. On the other hand, the minimal current symmetry subspace dominating current statistics for $|\lambda-\frac{\epsilon}{2}|<\frac{\epsilon}{2}$ is antisymmetric by pairs. This \emph{pair-fermionic transport regime} is again easily understood by noting that pairs of bulk qubits in singlet state are dark states of the dynamics (decoherence-free subspaces) which remain frozen in time and hence cannot accept excitations from the terminal qubits, effectively reducing the size of bulk and thus leading to a minimal current. In fact, this observation explains why $\mu(\lambda;N)$ does not depend on $N$ for $0<\lambda<\epsilon$, where the $N=2$ result always emerges. 

This severe dimensional reduction results from the symmetry of the $0<\lambda<\epsilon$ regime. In a similar way, we may now use the symmetry of the bosonic transport regime to strongly reduce the dimensionality of the total Hilbert space, hence allowing us to reach much larger network sizes than previously anticipated. In particular, a totally symmetric state of bulk qubits is univocally described by the total number of excitations in the bulk, $K\in[0,N-2]$, so the dimension of the total Hilbert space drops dramatically from an exponential $2^N$ to a linear $4(N-1)$ (see Appendix B for a detailed explanation {and refs. \cite{lipkin,ribeiro} for a similar dimensional reduction in the related Lipkin-Meshkov-Glick model}). Using this dimensional reduction, we were able to compute the LDF $\mu(\lambda;N)$ for quantum networks of size $N\leq 40$, see lines in Fig. \ref{fig3}, opening the door to a systematic study of finite-size effects in current statistics. For the LDF, our data strongly suggest a clear scaling $\mu(\lambda;N)=\mu(\lambda) + a(\lambda)N^{-1/2}$, see inset to Fig. \ref{fig3}, with $a(\lambda)$ some amplitude (note that $a(\lambda)=0$ for $0<\lambda<\epsilon$). This scaling yields an estimate of the LDF $\mu(\lambda)$ in the thermodynamic limit, see Fig. \ref{fig5}, confirming the presence of two clear kinks at $\lambda=0,\epsilon$. Notice that this LDF, as well as all finite-size LDFs in Fig. \ref{fig3}, obey the Gallavotti-Cohen fluctuation theorem $\mu(\lambda)=\mu(\epsilon-\lambda)$ as a result of microreversibility \cite{GC,ECM,LS,K,IFR,Gaspard,Agarwal,Mallick}. We also performed numerically the inverse Legendre transform of $\mu(\lambda)$ to obtain an estimate of the current LDF $G(q)$ in the $N\to\infty$ limit, see inset to Fig. \ref{fig5}. As expected, the kinks in $\mu(\lambda)$ translate into two current regimes, $|q|\in[|\la q_{\alpha_{\text{min}}} \ra|,|\la q_{\alpha_{\text{max}}} \ra|]$, where $G(q)$ is non-convex \cite{nonconvex} corresponding to a multimodal current distribution due to coexistence of different transport channels classified by symmetry.

\begin{figure}[t]
\includegraphics[width=8.5cm]{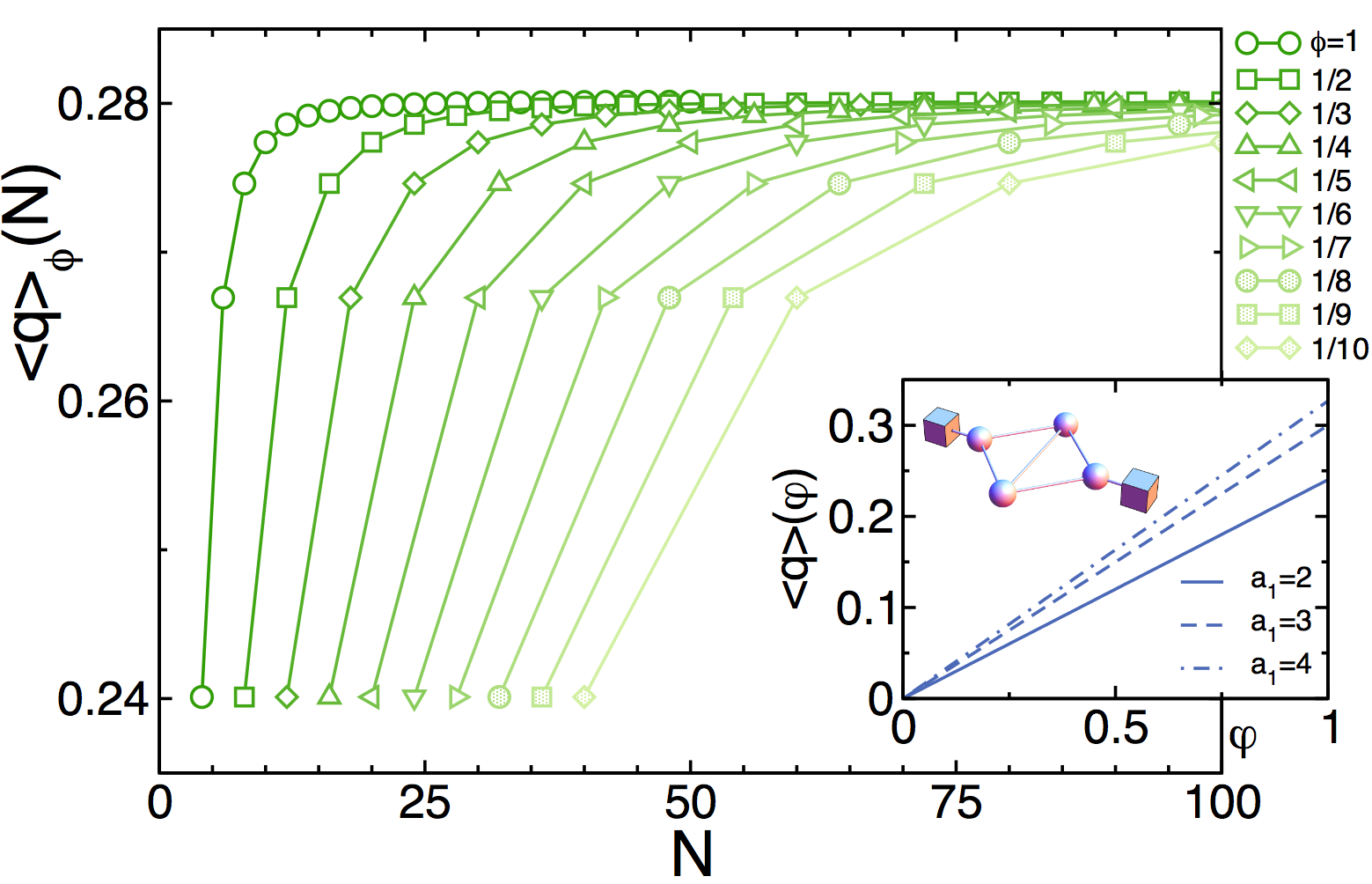}
\caption{\small (Color online) Size dependence of the average current $\la q\ra_{\phi,N}$ for a quantum network with an even number $(1-\phi)N$ of bulk qubits initialized in pair-antisymmetric states. The current increases both with $N$ and $\phi$, demonstrating symmetry-controlled transport. {System parameters as in Fig. \ref{fig3}.} {Inset: Average current for the sketched 4-qubit quantum thermal switch as a function of $\varphi$, the projection of the initial density matrix on the subspace of $\cBH$ corresponding to a totally-symmetric bulk, for different excitation pumping rates $a_1$ {(the other parameters as in Fig. \ref{fig3})}. This shows how the heat current between hot and cold reservoirs can be completely blocked, modulated or turned on by preparing the symmetry of the initial state.}
 }
\label{fig6}
\end{figure}

To illustrate the symmetry control over transport properties via initial state preparation, see Eq. (\ref{qmed}), we plot in the Fig. \ref{fig6} the average current as a function of the size of the quantum network for different initial states $\rho(0)$, prepared in a direct product configuration such that an even number $(1-\phi)N$ of bulk qubits are initialized in antisymmetric, singlet states by pairs, while the complementary set of bulk qubits are initially in a totally-symmetric state. As explained above, the antisymmetric pairs of qubits form dark states of the dynamics, remaining dynamically decoupled from the rest of the system. It is then easy to show (see Appendix B) that the resulting eigenvalue problem for $\cWl$, and consequently the average current and current statistics, thus correspond to those of a quantum network with $\phi N$ qubits and a totally-symmetric bulk. In this totally-symmetric (maximal current) setting a larger bulk means a larger current, so we expect the current to increase both with $N$ and $\phi$, as confirmed in Fig. \ref{fig6}. In this way, tuning the initialization parameter $\phi$ allows {to control} the average current for each $N$.

{The previous discussion suggests a modification of the network Hamiltonian in order to gain \emph{full} control of the heat current traversing the quantum system. In particular, by removing the interaction between the terminal qubits, it is possible to block completely the energy current which flows from the hot to the cold reservoir by initializing the bulk qubits in an antisymmetric-by-pairs state. This is most evident for the case of $N=4$ qubits, see Fig. \ref{fig1}.b. In fact, by initializing the system in an (otherwise arbitrary) mixed state such that the projection of the initial density matrix $\rho(0)$ on the symmetry eigenspace of $\cBH$ corresponding to a totally-symmetric bulk is fixed and equal to $\varphi\in[0,1]$, it is easy to show that the average current in this case is simply $\la q\ra=\varphi \la q_+\ra$, where $\la q_+\ra$ is the average current of the completely symmetric NESS $\rho_+^{\text{NESS}}$, see eq. (\ref{qmed}). Of course this is so because $\la q_-\ra=0$ due to the dynamical decoupling between terminal qubits produced by the frozen, dark state of the anti-symmetric bulk. As an example, the inset in Fig. \ref{fig6} shows the average current for the 4-qubit network in Fig. \ref{fig1}.b as a function of $\varphi$ for varying excitation pumping rates $a_1$. In this way, the combination of the simple network topology of Fig. \ref{fig1}.b with our symmetry results allows to design a \emph{symmetry-controlled quantum thermal switch}, where the heat current flowing between hot and cold reservoirs can be completely blocked, modulated or turned on by just preparing the symmetry of the initial state. Note that a \emph{nonlinear} control of the heat current can be also implemented by introducing a weighted interaction between terminal qubits.}

To end this section, we now study the effect of dephasing noise on the thermodynamics of currents, and in particular on the dynamic phase transitions and {spontaneous} symmetry-breaking phenomena discussed above. The interaction with a dephasing environment, that reduces the quantum coherent character of the system at hand, has been probed very important for the energy transfer in different nonequilibrium quantum networks, where noise-enhanced transport has been recently reported \cite{manzano:po13}. In order to simulate such environment we introduce a new set of Lindblad (dephasing) operators $L_m^{(\tiny \text{deph})}=\sqrt{\gamma} \sigma_m^+\sigma_m^-$, with $m\in [1,N]$, in the master equation (\ref{Lindblad}), which reduce the quantum coherences inside the system and, effectively, transforms the quantum transport in a classical one in a continuous way, depending on the dephasing parameter $\gamma$. As the dephasing Lindblad operators act locally on each qubit, they violate the bulk permutation symmetries of the original master equation. The new evolution equation hence mixes the original symmetry eigenspaces, thus leading to an unique NESS, independent of the initial steady state. In addition, the violation of the original strong symmetries immediately implies the disappearance of the twin dynamic phase transitions and the associated symmetry-breaking phenomenon at the fluctuating level, thus leading to a differentiable $\mu(\lambda)$ and a convex current LDF $G(q)$. Dashed thick lines in Fig. \ref{fig5} show $\mu(\lambda;N)$ as measured for systems with $N=2,4$ and 6 qubits and a dephasing parameter $\gamma=0.5$. In all cases, as expected, the LDF shows no kinks at $\lambda=0,\epsilon$ while obeys the Gallavotti-Cohen theorem for all currents. This result proves the essentially coherent character of the twin dynamic phase transitions and related symmetry-breaking phenomena, as they disappear whenever the bulk system dynamics is not purely coherent. A similar change of regime due to a dephasing channel has already been observed in lattices of qubits and harmonic oscillators far from equilibrium, where an arbitrary amount of dephasing makes the transport change from ballistic to diffusive \cite{manzano:pre12} (see also \cite{deph}).

\section{Discussion} 

We have shown in this paper how to harness symmetry to control transport and {current} statistics in open quantum systems. The action of different dissipative processes in the presence of a strong symmetry \cite{Prosen} drives a quantum systems to a degenerate steady state, which preserves part of the information present in the initial density matrix \cite{Albert}. By tailoring this information via initial state preparation, we are able to control both the average transport properties and the statistics of the current flowing through an open quantum systems. Remarkably, the coexistence of different transport channels at the heart of this control mechanisms {is associated with} a general dynamic phase transition in current statistics between two different symmetry-broken phases (maximal vs minimal current phases), which is accompanied by a twin dynamic phase transition for rare, reversed current fluctuations as a result of time-reversibility. This is reflected in non-analyticities and non-convex behavior in the current large deviation functions, which play a central role in nonequilibrium physics. {Motivated by the problem of energy harvesting and coherent transport in photosynthetic complexes}, we have applied these general results to study transport and current fluctuations in open quantum networks, finding excellent agreement with the predictions based on symmetry ideas. The experimental observation of the effects here described is accessible and desirable, as symmetry control of transport properties opens new avenues of future research worth exploring. The recent possibility of creating coherent cavity networks with complete connectivity \cite{elica} opens the door to potential  experimental realizations of these phenomena. In addition, these results call for symmetry-based design strategies for quantum devices with controllable transport properties. {In fact, our symmetry approach to transport has alowed us to introduce a \emph{symmetry-controlled quantum thermal switch}, i.e. a quantum qubit device where the heat current between hot and cold reservoirs can be completely blocked, modulated or turned on by playing with the initial state symmetry.} Note that a different transport control setup has been recently introduced by coupling vibrons to internal states of trapped ions in crystal lattices \cite{Qcontrol}. 

Dissipation has been typically considered negative for quantum information processing as it tends to destroy the coherent quantum effects which characterize the ultimate power of quantum computation. However, in a recent series of breakthroughs \cite{opensim,Kraus,Naturecomp,Naturecomp2,Wineland}, the situation has been reversed by carefully engineering the dissipation process to implement universal quantum computation \cite{Naturecomp} or in order to drive the open quantum system to desired (e.g. maximally entangled, matrix-product, etc.) states \cite{Wineland,Kraus,Naturecomp2}. Furthermore, controlled dissipation has been also used to \emph{protect} quantum states by extending their lifetime \cite{memo}. By combining these novel dissipation engineering techniques with design strategies based on symmetry principles, one can for instance create open quantum systems capable of storing \emph{at the same time} different coherent quantum states associated with the multiple, symmetry-protected steady states. We expect this line of research will trigger further advances in dissipative state engineering and dissipative quantum computation. 

From a general point of view, the results in this paper demonstrate the power of symmetry as a tool to obtain deep insights into nonequilibrium physics. This idea has been recently exploited to investigate nonequilibrium fluctuations in classical diffusive systems \cite{IFR}. By demanding invariance of the \emph{optimal path} responsible of a given fluctuation under symmetry transformations, a remarkable and general Isometric Fluctuation Relation (IFR) for current statistics was derived \cite{IFR} which links in a simple way the probability of different but isometric current fluctuations, and generalizes in this context the Gallavotti-Cohen fluctuation theorem. This new symmetry implies remarkable hierarchies of equations for the current cumulants and the nonlinear response coefficients which go far beyond Onsager's reciprocity relations and Green-Kubo formulas. The recent extension of large deviation formalism to open quantum systems \cite{Garrahan} allows to explore the quantum version of the IFR starting from the spectral properties of the deformed superoperator $\cWl$ and their behavior in the large size limit. The extension of the IFR to the quantum realm would open the door to further exact and general results valid arbitrarily far from equilibrium in a quantum setting, based on similar invariance principles.

Finally, in the open quantum network example studied above we have explored the role of \emph{geometrical} symmetries of the Hamiltonian on quantum transport, although our results apply to general symmetries. It would be interesting to find examples with other types of symmetries and systematic ways to implement symmetry control over transport properties. Clues are to be found in the recent application of symmetry principles to the problem of controllability and simulability of open quantum systems \cite{control}.

\phantom{xxx}

{\bf Acknowledgements.} Financial support from Spanish projects No. FIS2009-08451 (MICINN) and No. FIS2013-43201-P (MINECO), University of Granada, Junta de Andaluc\'{\i}a project P09-FQM4682, {GENIL PYR-2014-13 project} and Austrian Science Fund (FWF) F04012 is acknowledged.

\appendix

\section{Average current in a NESS}
\label{appA0}

We are interested in the average current for a generic Markovian open quantum system. As described in the main text, this average can be obtained from the moment generating function of the current distribution as
\be
\langle q \rangle=\lim_{t \to \infty} \frac{1}{t} \left[\partial_\lambda \ln Z_\lambda(t) \right]_{\lambda=0} \, , \nonumber
\ee
where $Z_\lambda(t)=\Tr[\rho_{\lambda}(t)]$. By differentiating the above expression taking into account the definition of $Z_\lambda(t)$, and noting that $\rho_\lambda(t)=\exp(+t\cWl)\rho(0)$, we have 
\be
\la q \ra = \lim_{t \to \infty} \left. \frac{\Tr[(\partial_\lambda \cWl) \rho_\lambda (t)]}{\Tr[\rho_\lambda (t)]} \right\vert_{\lambda=0} \, , 
\label{qq1}
\ee
where the new superoperator $\partial_\lambda \cWl$ is defined via
\be
(\partial_\lambda \cWl) \eta = \text{e}^{+\lambda} L_2 \eta {L_2}^\dagger - \text{e}^{-\lambda} L_1 \eta {L_1}^\dagger  \, , \quad \forall \eta \in \cBH \, ,
\ee
as derived from the definition of $\cWl$ in Eq. (2) of the main text. If we now restrict the initial density matrix to a particular symmetry subspace, $\rho(0)\in \cBaa$, we have that $\lim_{t\to \infty} \rho_\lambda(t)\vert_{\lambda=0}=\rho_{\alpha}^{\scriptscriptstyle\text{NESS}}$, which is normalized, $\Tr[\rho_{\alpha}^{\scriptscriptstyle\text{NESS}}]=1$, and therefore
\be
\la q_{\alpha}\ra = \Tr[L_2\rho_{\alpha}^{\scriptscriptstyle\text{NESS}}L_2^\dagger]-\Tr[L_1\rho_{\alpha}^{\scriptscriptstyle\text{NESS}}L_1^\dagger] \, . 
\label{qq2}
\ee
On the other hand, for a general $\rho(0)\in \cBH$ we may use in Eq. (\ref{qq1}) the spectral decomposition 
\be
\rho_\lambda(t)=\sum_{\alpha\beta\nu} \text{e}^{+t\mu_\nu(\lambda)} \left(\tilde{\omega}_{\alpha\beta\nu}(\lambda),\rho(0)\right)\omega_{\alpha\beta\nu}(\lambda) \, , \nonumber
\ee
with $\mu_\nu(\lambda)$ and $\omega_{\alpha\beta\nu}(\lambda)$ the eigenvalues and associated (right) eigenfunctions of $\cWl$, see main text. As for $\cWl$, the new superoperator $\partial_\lambda \cWl$ leaves invariant the symmetry subspaces, $(\partial_\lambda \cWl) \cBa \subset \cBa$, so $\Tr[(\partial_\lambda \cWl) \rho_\lambda (t)]=\sum_{\alpha\alpha\nu} \text{e}^{+t\mu_\nu(\lambda)} \left(\tilde{\omega}_{\alpha\alpha\nu}(\lambda),\rho(0)\right) \Tr[(\partial_\lambda \cWl)\omega_{\alpha\alpha\nu}(\lambda)]$. By noting that for $\lambda=0$ the largest eigenvalue of $\cWl$ within each symmetry eigenspace $\cBaa$ is necessarily 0, with associated normalized right eigenfunction $\omega_{\alpha\alpha 0}(\lambda=0)=\rho_{\alpha}^{\scriptscriptstyle\text{NESS}}$ and dual $\tilde{\rho}_{\alpha}^{\scriptscriptstyle\text{NESS}}$, we hence obtain
\be
\la q \ra = \frac{\sum_{\alpha} \la q_{\alpha}\ra (\tilde{\rho}_{\alpha}^{\scriptscriptstyle\text{NESS}},\rho(0))}{\sum_{\alpha} (\tilde{\rho}_{\alpha}^{\scriptscriptstyle\text{NESS}},\rho(0))} \, . \nonumber
\ee

\section{Dimensional reduction of the Hilbert space for the open quantum network}
\label{sec:appendixa}

As briefly described in the main text, we may use the totally symmetric nature of the maximal current fluctuating phase, $|q|>|q_{\alpha_{\text{max}}}|$, to drastically reduce the dimension of the relevant Hilbert space in this regime. In this way the dimension of the problem for a network of $N$ qubits can be reduced from an exponential $O(2^N)$ to a linear scaling $O(N)$. {Such dimensional reduction was already noted in previous studies of the related Lipkin-Meshkov-Glick model \cite{lipkin,ribeiro}.} Combining this result with the size-independence found for the pair-antisymmetric, minimal current phase, $|q|<|q_{\alpha_{\text{min}}}|$, this technique allows us to reach network sizes up to $N=40$ qubits, much larger than what any numerical method can handle with general multipartite qubit states. This size range is enough to study the dominant scaling for finite-size corrections, thus allowing us to obtain estimates of the cumulant generating function of the current distribution and the current large deviation function in the thermodynamic limit, see Fig. \ref{fig5} in the main text.

\subsection{Totally symmetric regime ($|q|>|q_{\alpha_{\text{max}}}|$)}

We start by noting that a completely symmetric state of bulk qubits is univocally described by the total number of excitations in the bulk. Let $\ket{\mathbf{n}}\equiv \otimes_{i=1}^N \ket{n_i}\in \cH$, with $\ket{n_i}=\ket{0}$ or $\ket{1}$, be a state of the Hilbert space for our open quantum network as expressed in the computational basis, and denote as $N_{\text{b}}\equiv N-2$ the number of bulk qubits. An arbitrary state with a totally symmetric bulk can be thus written as
\be
\ket{K;n_1,n_N} = \frac{1}{\sqrt{N_{\text{b}} \choose K}}\sum_{n_2\ldots n_{N-1}=0,1} \ket{\mathbf{n}} \, \delta \Big(K-\sum_{i=2}^{N-1} n_i \Big) \, 
\label{nsim}
\ee
where $K\in[0,N_{\text{b}}]$ is the total number of excitations in the bulk in this symmetric state, and the ${N_{\text{b}} \choose K}$ in the normalization constant counts the number of ways of distributing $K$ excitations among $N_{\text{b}}$ bulk qubits. The dichotomy between bulk and terminal qubits allows us to decompose the Hamiltonian (\ref{hamil}) of the qubit network as $H = H_0 + H_{\text{b}} + H_{\text{I}}$, where
\ben
&H_0& \equiv h \sum_{i=1}^N \sigma_i^+ \sigma_i^- \, , \label{H0} \\
&H_{\text{b}}& \equiv J \sum_{\substack{i=2 \\ i<j\le N-1}}^{N-2} \Delta_{ij} \, , \label{Hb}
\een
with the definition $\Delta_{ij} \equiv (\sigma_i^+\sigma_j^- + \sigma_i^-\sigma_j^+)$, and
\be
H_{\text{I}} = J  \Big[ (\sigma_1^+ + \sigma_N^+)\Delta_- + (\sigma_1^- + \sigma_N^-)\Delta_+ + \Delta_{1N}\Big] \, ,
\label{Hint}
\ee
where we further define $\Delta_{\pm} \equiv \sum_{i=2}^{N-1} \sigma_i^{\pm}$. It is then trivial to show that the on-site contribution to the Hamiltonian, $H_0$, is diagonal in the lower-dimensional basis defined by the states (\ref{nsim}), i.e. $H_0 \ket{K;n_1,n_N} = h (K+n_1+n_N) \ket{K;n_1,n_N}$, so we can write
\be
H_0 = h \sum_{\substack{K=0 \\ n_1,n_N=0,1}}^{N_{\text{b}}} (K+n_1+n_N)  \ket{K;n_1,n_N} \bra{K;n_1,n_N}
\label{aH0}
\ee
To understand the action of the bulk self-interaction part $H_{\text{b}}$ on states (\ref{nsim}), first notice that the operators $\Delta_{ij}$ simply exchange the states of qubits $i$ and $j$ whenever they are different, yielding zero otherwise, i.e. $\Delta_{ij} \ket{\mathbf{n}} = \delta_{n_i,1-n_j}\ket{\mathbf{n}}_{ij}$, where $\ket{\mathbf{n}}_{ij}$ is the state in the computational basis resulting from exchanging $n_i \leftrightarrow n_j$ in $\ket{\mathbf{n}}$. Using this expression when operating with $H_{\text{b}}$ on the bulk-symmetric states (\ref{nsim}), it is easy to show that $H_{\text{b}}$ is also diagonal in the basis defined by $\ket{K;n_1,n_N}$, with a prefactor counting the number of distinct pairs that we can form with $K$ $\ket{1}$'s and $(N_{\text{b}}-K)$ $\ket{0}$'s, so 
\be
H_{\text{b}} = J \sum_{\substack{K=0 \\ n_1,n_N=0,1}}^{N_{\text{b}}} K (N_{\text{b}}-K) \ket{K;n_1,n_N} \bra{K;n_1,n_N} \, .
\label{aHb}
\ee
It is now straightforward to show that the operators $\Delta_{\pm}$ move the state $\ket{K;n_1,n_N}$ to $\ket{K\pm 1;n_1,n_N}$, with a prefactor that counts the number of ways of distributing the pertinent excitations among $(N_{\text{b}}-1)$ bulk sites and takes into account the different normalizations. In particular, $\Delta_{\pm}\ket{K;n_1,n_N} = D_K^{\pm} \ket{K\pm 1;n_1,n_N}$, with 
\ben
D_K^+ &=& \sqrt{(K+1)(N_{\text{b}}-K)} \, , \label{dp} \\
D_K^- &=& \sqrt{K(N_{\text{b}}-K+1)} \, , \label{dm}
\een
so we may write
\be
\Delta_{\pm} = \sum_{\substack{K=k_{\pm} \\ n_1,n_N=0,1}}^{N_{\text{b}}-(1-k_{\pm})} D_K^{\pm} \ket{K\pm 1;n_1,n_N} \bra{K;n_1,n_N} \, ,
\label{aDpm}
\ee
with $k_{\pm}\equiv (1\mp 1)/2$. In this way the Hamiltonian of the open quantum network with a completely symmetric bulk can be fully written in terms of the low-dimensional basis formed by vectors (\ref{nsim}). As the Lindblad operators in the master equation (\ref{Lindblad}) only act on the network terminal qubits, 
the dimension of the problem in the totally symmetric regime is reduced spectacularly from the original $2^N$ to a much lower dimension $4(N-1)$, which scales linearly with the number of qubits.

\subsection{Pair-antisymmetric regime ($|q|<|q_{\alpha_{\text{min}}}|$)}

In this section we want to show that, for an open and fully-connected quantum network of size $N$ with a pair of bulk qubits in antisymmetric state, the associated deformed Lindblad superoperator  (\ref{rhol}) --and the corresponding eigenvalue problem-- is equivalent to the superoperator obtained for a network with $N-2$ qubits. This stems from the antisymmetric pair of qubits forming a dark state of the dynamics, which remains frozen in time and effectively decouples from the rest of the system. 

We hence start with a network with $N$ qubits, such that the pair formed by the (otherwise arbitrary) bulk qubits $a$ and $b$ is in an antisymmetric state. This means that our initial density matrix can be written as $\rho_-\equiv \ket{-}\bra{-}_{ab} \otimes \rho_{N-2}$, where $\ket{-}=\frac{1}{\sqrt{2}}(\ket{10}-\ket{01})$ is the singlet, antisymmetric state, and $\rho_{N-2}$ is an arbitrary reduced density matrix for the remaining $N-2$ qubits. To see how the deformed Lindblad superoperator (\ref{rhol}) acts on this pair-antisymmetric mixed state, we first decompose the Hamiltonian (\ref{hamil}) in three parts, $H=H_{ab} + H_{N-2} + H_{\text{int}}$ with
\ben
H_{ab} &=& h(\sigma_a^+ \sigma_a^- + \sigma_b^+ \sigma_b^-) + J \Delta_{ab} \, , \nonumber \\
 H_{\text{int}} &=& J \Big[ (\sigma_a^+ + \sigma_b^+) \sum_{\substack{k=1 \\ k\neq a,b}}^N \sigma_k^- +  (\sigma_a^- + \sigma_b^-) \sum_{\substack{k=1 \\ k\neq a,b}}^N \sigma_k^+ \Big]  \, , \nonumber
\een
and $H_{N-2}$ is the Hamiltonian (\ref{hamil}) for $N-2$ qubits excluding qubits $a$ and $b$. It is now a simple task to show that the terms $H_{ab}$ and $H_{\text{int}}$ of the Hamiltonian decomposition above commute with any pair-antisymmetric density matrix of the form $\rho_-$, $[H_{ab},\rho_-]=0=[H_{\text{int}},\rho_-]$, and hence the eigenvalue problem boils down to that of a network with $(N-2)$ qubits. In particular,
\ben
\dot{\rho}_- &=& -\ii [H,\rho_-] + {\cal L}_1^{(\lambda)} \rho_- + {\cal L}_N \rho_- = \cWl^{(N)} \rho_- \, \nonumber \\
&=& \ket{-}\bra{-}_{ab} \otimes \Big( -\ii [H_{N-2},\rho_{N-2}] + {\cal L}_1^{(\lambda)} \rho_{N-2} + {\cal L}_N \rho_{N-2} \Big) \nonumber \\
&=& \ket{-}\bra{-}_{ab} \otimes \Big(\cWl^{(N-2)} \rho_{N-2}\Big) = \ket{-}\bra{-}_{ab} \otimes \dot{\rho}_{N-2} \, , \nonumber
\een
where ${\cal L}_1^{(\lambda)}$ and ${\cal L}_N$ are the Lindblad superoperators which can be defined from Eq. (\ref{rhol}) above. Interestingly, using this method in a recursive manner it can be proved that the eigenvalue problem for any open quantum network of arbitrary size with a pair-antisymmetric bulk (i.e. with non-overlapping pairs of bulk qubits in singlet state) can be reduced to the case $N=2,3$, depending on $N$ beign even or odd. This explains why the pair-antisymmetric current fluctuation regime $|q|<|q_{\alpha_{\text{min}}}|$ in our quantum network is size-independent, providing a dramatic dimensional reduction of the relevant Hilbert space.


\begin{thebibliography}{99}
\bibitem{note1}  No positive eigenvalues for $\cW$ are possible due to trace preservation. Moreover, we are implicitly assuming here that no oscillating phases are present in the null subspace of the Lindblad superoperator (see ref. \cite{Albert} for more details on these exotic phases). This assumption implies in turn that the eigenvalue of the deformed superoperator $\cWl$ with largest real part is just real.
\bibitem{nonconvex} Technically speaking, the current LDF $G(q)$ can be either non-convex or affine in the intervals $|q|\in[|\la q_{\alpha_{\text{min}}} \ra|,|\la q_{\alpha_{\text{max}}} \ra|]$, and this cannot be inferred from $\mu(\lambda)$ \cite{Touchette}. However, a current distribution with a completely flat regime between $\la q_{\alpha_{\text{min}}} \ra$ and $\la q_{\alpha_{\text{max}}} \ra$ is a highly unlikely physical situation, so we expect generically a non-convex $G(q)$, corresponding to a multimodal current distribution reflecting the different, coexisting transport channels present in an open quantum system with a strong symmetry.
\bibitem{kink} Note that $\mu(\lambda)$ in equilibrium might still exhibit a (symmetric) kink at $\lambda=0$ due to some pathological behavior of current fluctuations in the dominant symmetry subspace in equilibrium, as e.g. a symmetric double-hump $G(q)$. This potential kink would be however unrelated to the underlying symmetry of the open quantum system.
\end{thebibliography}
\end{document}